\documentclass[prd,a4paper,twocolumn,nofootinbib]{revtex4}
\usepackage{mathrsfs}
\usepackage{graphicx}
\usepackage{latexsym}
\usepackage{amsmath}
\usepackage{amssymb}
\usepackage{latexsym}
\usepackage{textcomp}
\usepackage{amsbsy}
\usepackage{graphics}
\usepackage{slashed}
\usepackage{braket}
\usepackage{xcolor}
\usepackage{calc}
\usepackage{enumitem}
\newlength{\depthofsumsign}
\begin{document}
\title{Mass gap in the weak coupling limit of $(2+1)$ SU(2) lattice gauge theory}
\author{Ramesh Anishetty} 
\email{ramesha@imsc.res.in}
\author{T. P. Sreeraj}  \email{sreerajtp@imsc.res.in}
\affiliation{The Institute of Mathematical Sciences, C.I.T campus, Taramani, Chennai, India} 
\begin{abstract}
We develop the dual description of $2+1$ SU(2) lattice gauge theory as interacting `abelian like' electric loops by using Schwinger bosons. ``Point splitting" of the lattice enables us to construct explicit Hilbert space for the gauge invariant theory which in turn makes dynamics more transparent. Using path integral representation in phase space, the interacting closed loop dynamics is analyzed in the weak coupling limit to get the mass gap.
\end{abstract} 

\maketitle

\section{Introduction}
\label{sec:intro} 
There had been many attempts in the past to reformulate gauge theories in terms of gauge invariant Wilson loops \cite{wilsonloop,loop2,ms,loop3} both in the continuum as well as on the lattice. 
Since all Wilson loops are not independent even on a finite lattice, such a description based on non-local Wilson loops is over-complete. Constructing a complete algebra of observables in terms of loops and a complete loop basis for gauge theory Hilbert space is a non-trivial \cite{loop2,loop3,ms} task. 
Loop description of gauge theories on a lattice has been shown to be equivalent to a description based on integer local quantum numbers in dual space satisfying triangle inequalities \cite{ani} at each site. This is most conveniently realized in a Hamiltonian formulation using 
  the prepotential representation of gauge theories on a square lattice.
  The matrix elements of the magnetic part of the Hamiltonian in such a basis involves higher Wigner coefficients \cite{prep} and are complicated to analyze. 
  
  In this paper, we construct a complete, gauge invariant, local description of SU(2) lattice gauge theory (LGT) in $2+1$ dimensions by solving explicitly the gauge constraint on the 'electric' configuration space which is then further simplified by envisaging what is called a 'point split lattice' (see section \ref{pssec}). Further more, a manifest gauge invariant Hilbert space is constructed at every site on the split lattice.
  The local gauge invariant basis thus created is parameterized in terms of  three quantum numbers at each site of the new lattice which are independent  and have to satisfy triangle inequalities. These quantum numbers gives a measure of the gauge invariant electric flux lying on the links of the lattice. Therefore, gauge theory reduces to a theory of interacting electric flux loops. Matrix elements of Hamiltonian are simpler to analyze in this basis especially in the weak coupling limit.
   Such a local, complete basis allows us to construct a gauge invariant path integral for SU(2) LGT in the phase space.  This when analyzed in the weak coupling limit, 
    the mean gauge invariant electric flux becomes large and the small spatial electric flux loops dominate in the vacuum state. Further, the triangle inequality constraints, which are 
   very difficult to solve in general, becomes sub-dominant in the weak coupling limit.

The plan of the paper is as follows:
In section \ref{shf}, we give a brief review of the Kogut-Susskind Hamiltonian formulation \cite{ks} on a lattice. In section \ref{pssec}, we describe the procedure of modifying the lattice which we call 'point splitting'. Further, we describe the prepotential representation of SU(2) lattice gauge theory. In section \ref{basissec}, the construction of a complete, gauge invariant local basis on the modified (point split) lattice is described. The Hilbert space of the point split lattice is described by three gauge invariant quantum numbers $n_1,n_2,m$ per site. In section \ref{hamps}, the Hamiltonian on the modified lattice is given in terms of the prepotential operators. In section \ref{pisec}, we develop a gauge invariant path integral on the lattice by defining phases conjugate to $n_1,n_2,m$ and using the usual time slicing method. 
 A weak coupling expansion is developed in section \ref{weakc}. Normalization factors of the new basis is derived in appendix \ref{app:on}. The details of the construction of the path integral and the weak coupling expansion are given in appendix \ref{app:pi} and \ref{wc} respectively.
\section{Hamiltonian formulation of SU(2) lattice gauge theory}
\label{shf} 
\begin{figure}[b]
	\centering
	\includegraphics[scale=.7]{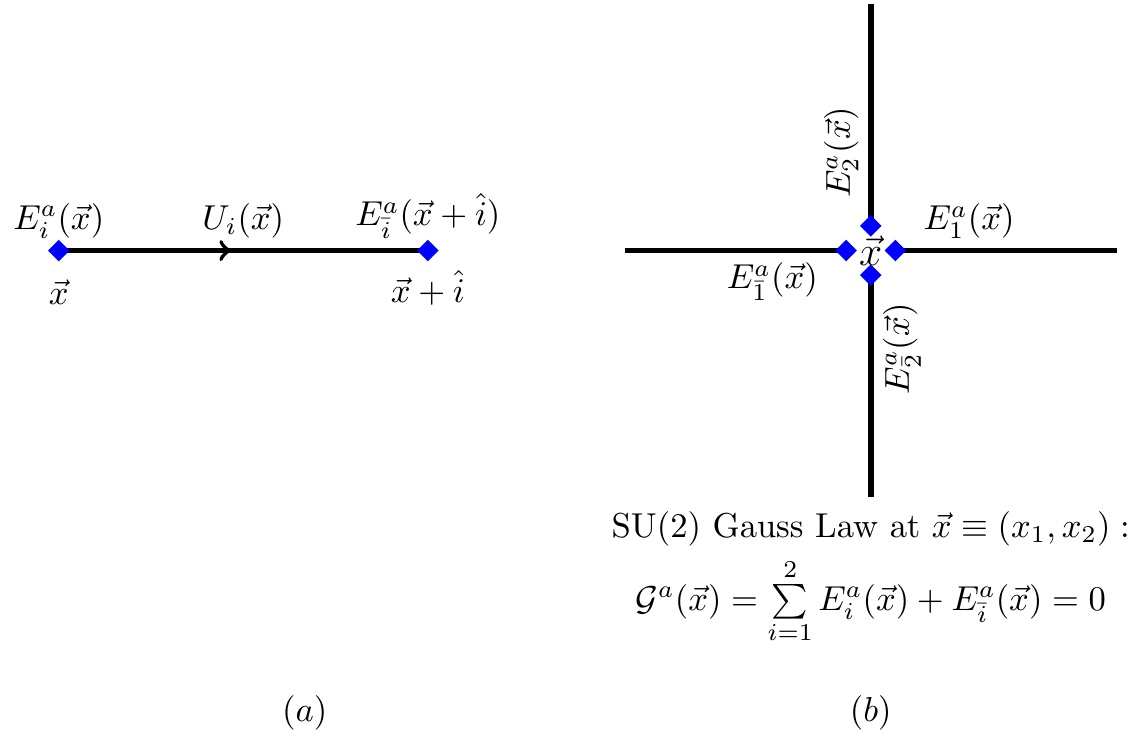}
	\caption{(a) A general link operator $U_i(\vec{x})$ and the corresponding left and right electric fields $E_i(\vec{x})$ and $E_{\bar i}(\vec{x}+i)$. (b) Gauss law at site $\vec{x}$.}
	\label{link} 
\end{figure}

In this section, we review the essential features of Kogut-Susskind formulation \cite{ks} of SU(2) lattice gauge theory. We consider a square lattice.
The basic variables of Kogut Susskind Hamiltonian formulation \cite{ks} are the SU(2) link operators $U_i(\vec{x})$ 
 and the conjugate left and right electric fields [see figure \ref{link}] $E_i(\vec{x})$ and $E_{\bar i}(\vec{x}+i)$. $U_i(\vec x)$ is an operator valued $2 \times 2$ SU(2) matrix lying on the link starting at site $\vec{x}$ along the $i$ direction The electric fields are SU(2) lie algebra elements and can be expanded as $E_{i/{\bar i}}(\vec{x})= E_{i/{\bar i}}^a(\vec{x}) \frac{\sigma^a}{2}$.
They satisfy the following commutation relations:
\begin{align}
\label{ccr} 
\left[E^{a}_{i}(\vec{x}),U_i(\vec{x})\right] & =   - \left(\frac{\sigma^a}{2}~ U_i(\vec{x})\right)  \\ 
\left[E^{a}_{\bar i}(\vec{x}+\hat i),U_i(\vec{x})\right] & =  ~~\left(U_i(\vec{x})~\frac{\sigma^a}{2}\right) \nonumber  
\end{align} 
$\sigma^{a}$ are the Pauli matrices.
The different components of left and right electric fields satisfy the SU(2) algebra and the left electric field commutes with the right electric field.
$E_i^a(\vec{x})$ and $E_{\bar i}^a(\vec{x}+i)$ generate left and right SU(2) transformations on $U_i(\vec{x})$ and are related as follows: 
\begin{align}
E^a_{\bar i}(\vec{x}+i)=R_{ab}(U) E^a_{i}(\vec{x}) 
\label{eler}
\end{align}
Above, $R_{ab}(U) =\frac{1}{2}Tr\big(U\sigma^aU^\dagger\sigma^b\big)$. Therefore,
\begin{align}
E_{i}^a(\vec{x})E_{i}^a(\vec{x})&=E_{\bar i}^a(\vec{x}+\hat{i})E_{\bar i}^a(\vec{x}+\hat{i})=E^2(\vec{x})
\label{u1}
\end{align} 
We call the above relation `link constraint'. The lattice electric fields are dimensionless and are related to their dimensionful continuum counterparts by powers of the lattice spacing $a$. Therefore, in the continuum the usual mutually commuting electric field components are recovered.
Under a local gauge transformation $\Lambda(\vec{x})$, the link operator and the electric fields transform as follows: 
\begin{align} 
U_i(\vec{x})&\xrightarrow{\Lambda}{} \Lambda(\vec{x}) U_i\Lambda^\dagger(\vec{x}+\hat{i})\nonumber\\
E_{i}(\vec{x})&\xrightarrow{\Lambda}{} \Lambda(\vec{x})E_{i}(\vec{x})\Lambda^\dagger(\vec{x})\\
E_{\bar i}(\vec{x}+\hat{i})&\xrightarrow{\Lambda}{} \Lambda(\vec{x}+\hat{i})E_{\bar i}(\vec{x}+\hat{i})\Lambda^\dagger(\vec{x}+\hat{i})\nonumber
\end{align}
The Hamiltonian is given by: 
\begin{align}
\label{ksham}
H=\frac{\tilde{g}^2}{2}\sum\limits_{\vec{x},i}E^a_i(\vec{x})E^a_i(\vec{x})+\frac{1}{2g^2}\sum\limits_{p} \big[2-TrU_p\big]
\end{align}
In (\ref{ksham}), $p$ denotes plaquettes. We study the general case where the electric coupling is ${\tilde g}$ and magnetic coupling is $g$.  Since, Hamiltonian has mass dimension of 1 in the natural units ($\hbar=c=1$) and $E^a$ and $U$ are dimensionless it follows \cite{loop3} from naive continuum limit that $\tilde{g}^2 \sim \frac{1}{a}$ and $g^2 \sim a$, where $a$ is the lattice constant.
 Above Hamiltonian along with the following Gauss law constraints (\ref{gl}) (see figure \ref{link}) completely defines the theory: 
\begin{align}
\label{gl}
{\cal G}^a(\vec{x})=
\sum\limits_{i=1}^{2} \Big[E_{i}^a(\vec{x})+E_{\bar i}^a(\vec{x})\Big]=0
\end{align}
${\cal G}^a(\vec{x})$ is the generator of gauge transformations at site ${\vec x}$.  

A complete local electric basis can be constructed \cite{ani} at each site by quantum numbers of  $(E_1)^2~,~ (E_{\bar 1})^2~,~(E_2)^2~,~(E_{\bar 2})^2~,~(E_{\bar 1}+E_1)^2~,~(E_{\bar 2}+E_2)^2~,~(E_{\bar 1}+E_1+E_{\bar 2}+E_2)^2~,~(E_{\bar 1}+E_1+E_{\bar 2}+E_2)^{(3)}$.
Above, $(3)$ in the superscript denotes the third component. The Gauss law constraint (\ref{gl}) requires that the last two operators of the set of operators above vanish in the physical Hilbert space. This along with eqn.(\ref{u1}) implies that 
a complete gauge invariant basis at each site is given by $|j_i,j_{\bar i},j_{i\bar{i}}\rangle$, where $j_i,j_{\bar i},j_{i\bar{i}}$ are quantum numbers labelling the eigenstates of $(E_i)^2$,$(E_{\bar i})^2$ and $(E_1+E_{\bar 1})^2=(E_2+E_{\bar 2})^2$, for the choice given in figure \ref{ps}(c).
This choice is not unique, there are two other such addition schemes possible leading to three different basis at each site related by unitary transformations. These three schemes are graphically represented in figure \ref{ps}.
\section{Point splitting and prepotential representation}
\label{pssec}
\begin{figure}
	\centering
	\includegraphics[scale=.85]{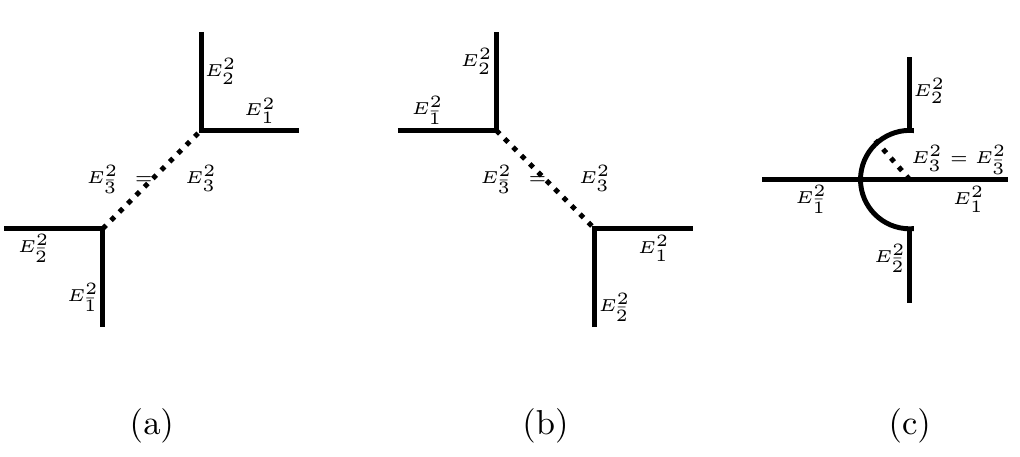}
	\caption{The 3 possible SU(2) addition schemes which can be used in the construction of  a gauge invariant basis at each site. New links are introduced along the dotted lines to make the scheme manifest. Introduction of  new links along with 2 new Gauss laws constraints at the sites where dotted and solid lines meet and a link constraint for the new link electric fields keeps physics unchanged. We choose scheme (c) in this paper.} 
		\label{ps}
\end{figure}
\begin{figure*}
	\includegraphics[]{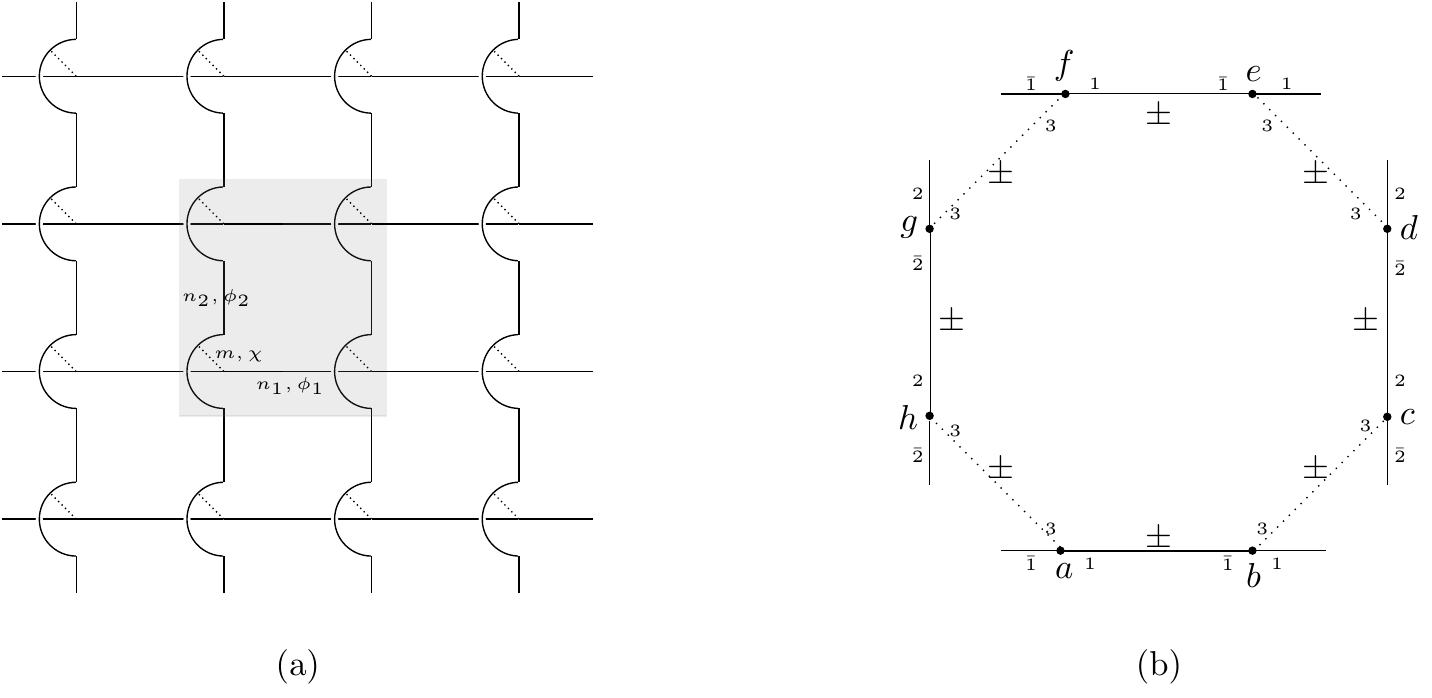}
	\caption{ (a). The point split lattice (ps-lattice) obtained after performing the splitting of point at each site. (b) Plaquette becomes an octagon after point splitting. Action of $TrU_o$ increases $n_i/m$ on each link on the octagon by $\pm 1$. }
	\label{pslattice}
\end{figure*}
Out of the three quantum numbers at each site, two gives a measure of the electric flux on the links and the third gives a measure of the coupled electric flux according to the SU(2) addition scheme chosen. In this paper, we use the addition scheme $(c)$ in figure \ref{ps}. We now introduce new links and corresponding link operators and electric fields (and corresponding link constraints) along the dotted lines (see figure \ref{ps}),
  This way, 
the angular momentum addition scheme used to solve the Gauss law becomes manifest. 
Now, each 4-vertex (i.e, sites where 4 links meet) is converted to two 3-vertices. Each of them are labeled by the same position index as the original 4-vertex. Therefore, the link operator on the dotted line connecting two 3-vertices labelled by $\vec{x}$ is denoted as $U_3(\vec{x})$ and the corresponding left and right electric fields are denoted as $E_3(\vec{x})$ and $E_{\bar 3}(\vec{x})$. Each 3-vertex is associated with the Gauss law:
\begin{align}
E_1(\vec{x})+E_{\bar 1}(\vec{x})+ E_{3}(\vec{x})=0\nonumber \\
E_2(\vec{x})+E_{\bar 2}(\vec{x})+ E_{\bar{3}}(\vec{x})=0
\end{align} 
 Lattice becomes a collection of 3-vertices connected by links (see figure \ref{pslattice}).
We call this process `point splitting' and the new lattice after point splitting a ps-lattice. 

The ps-lattice describes the same physics as the original lattice. This equivalence can be seen as follows. Without any loss of generality, Gauss law at one of the two new 3-vertex can be used to gauge fix the new link operator (along dotted line) to identity. This implies that (by eqn.(\ref{eler})) after gauge fixing $E_i(\vec{x})=-E_{\bar i}(\vec{x})$ on the new link. Therefore, we get the old Gauss law involving 4 electric fields back, hence we are back to the old lattice. 
The degrees of freedom per site are easily counted on a ps-lattice to be three as expected. 
 In this paper, we choose the splitting scheme (c) in figure. \ref{ps} at each site of the lattice. 
 This leads to a ps-lattice given in figure \ref{pslattice}. The $TrU_p$ in the Hamiltonian becomes the trace of product of 8 link operators around an octagon (see figure \ref{pslattice}(b)):
\[ TrU_p \rightarrow TrU_o.\] This can be easily shown to be true as $TrU_o$ reduces to $TrU_p$ by the gauge fixing discussed above. Only the electric fields at the  solid (old x-links and y-links) links contribute to the Hamiltonian.
\subsection{Prepotential representation of SU(2) LGT}
The gauge invariant basis on the ps-lattice can be better analyzed in the prepotential \cite{rmi,prep,rmi2} representation. Therefore, we briefly review the prepotential representation of SU(2) LGT. 
SU(2) lattice gauge theory can be reformulated in terms of harmonic oscillator doublets called Schwinger bosons. Left and right electric fields are written in terms of these 
Schwinger bosons as follows: 
\begin{align}
\label{prepe}
E_i^{a}(\vec{x}) &\equiv 
a^{\dagger}_{i,\alpha}(\vec{x})\left(\frac{\sigma^{a}}{2}\right)_{\alpha \beta}a_{i,\beta}(\vec{x}), \nonumber\\ 
E_{\bar i}^{a}(\vec{x}+\hat{i}) &\equiv 
a^{\dagger}_{{\bar i},\alpha}(\vec{x}+\hat{i})\left(\frac{\sigma^{a}}{2}\right)_{\alpha\beta}a_{\bar{i},\beta}(\vec{x}+\hat{i}).
\end{align}
Above, $a^{\dagger}_{i,\alpha}(\vec{x})$ and $a_{{\bar i},\alpha}^{\dagger}(\vec{x}+\hat{i})$ are the harmonic oscillator doublets at the left and right of the link operator $U_{i}(\vec{x})$. Similar prepotential representation can be made for the left and right electric fields of the dotted link also by putting $i=3$ and replacing $x+\hat{i}$ with $\vec{x}$. 
The link constraint (\ref{u1}) when written in terms of Schwinger bosons read: 
\begin{align}
\label{u1prep}
 a_i^\dagger(\vec{x})\cdot a_i(\vec{x})=a_{\bar i}^\dagger(\vec{x}+\hat{i})\cdot a_{\bar i}(\vec{x}+\hat{i})\equiv \hat{N}
 \end{align}
The number of Schwinger bosons at the left end of a link equals that at the right end. The link operator can be written in terms of Schwinger bosons as : 
{\footnotesize
\begin{align}
\label{prepu}
(U_{i})_{{}_{\alpha\beta}}(\vec{x}) = f(\hat{N}) \Big[\tilde{a}^\dagger_{i,\alpha}(\vec{x}) a^\dagger_{\bar{i},\beta}(\vec{x}+\hat{i})+a_{i,\alpha}(\vec{x})\tilde{a}_{{\bar i},\beta}(\vec{x}+\hat{i})\Big]f(\hat{N})
\end{align}}
Above, $\tilde{a}^\dagger=\epsilon_{\alpha\gamma}a^\dagger_\gamma$ and $f(\hat{N})=\frac{1}{\sqrt{\hat{N}+1}}$.  
$U_{\alpha\beta}$ changes the number of Schwinger bosons on the link by $\pm1$.
 Since, $E^2=\frac{\hat{N}}{2}(\frac{\hat{N}}{2}+1)$, the electric part of the Hamiltonian counts the number of Schwinger bosons while the magnetic part ($TrU_o$) fluctuates the number.
 Schwinger bosons at any 3-vertex transform as  
\begin{align}
a_\alpha \xrightarrow{\Lambda} \Lambda_{\alpha\beta} ~a_\beta\nonumber
\end{align}
where $\Lambda$ is a gauge transformation at the 3-vertex. 
Note that $\tilde{a}^\dagger_\alpha$ transforms the same way as $a_\alpha$.
\begin{figure}[b]
	\includegraphics[]{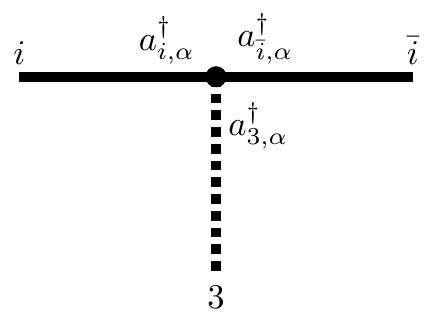}
	\caption{ A basic 3 vertex on the PS-lattice. Three prepotentials associated with the 3-vertex is also shown. The three links are labelled by $i, \bar{i}, 3$ }
	\label{3vertex}
\end{figure}
\subsection{Gauge invariant local basis on the ps-lattice}
\label{basissec}
ps-lattice is a collection of  3-vertices connected together by links.
At any 3-vertex, there are 3 prepotentials(see figure \ref{3vertex}) denoted as $a_{i,\alpha}^\dagger,a_{\bar{i},\alpha}^\dagger,a_{3,\alpha}^\dagger$ where $ i=1,2$ according to whether the vertex contains x-links or y-links. On every 3-vertex, $(a^\dagger_{i}\cdot \tilde{a}^\dagger_{\bar i}), (a^\dagger_{\bar i} \cdot \tilde{a}^\dagger_3), (a^\dagger_3 \cdot \tilde{a}^\dagger_i)$ are locally gauge invariant. So, a complete, orthonormal, gauge invariant local basis at a site is given by: 
\begin{align}
|l_{i\bar{i}},l_{\bar{i}3},l_{3i}\rangle= \frac{(a_i^\dagger \cdot \tilde{a}_{\bar i}^\dagger)^{l_{i\bar{i}}} (a^\dagger_{\bar i} \cdot \tilde{a}^\dagger_3)^{l_{\bar{i}3}} (a^\dagger_3 \cdot \tilde{a}^\dagger_i)^{l_{3i}}~}{\sqrt{(l_{i\bar{i}}+l_{i3}+l_{\bar{i}3}+1)!(l_{i\bar{i}})!(l_{3i})!(l_{\bar{i}3})!}} ~ |0\rangle
\label{psbasis}
\end{align}
In (\ref{psbasis}), $|0\rangle$ is the strong coupling vacuum : $a_\alpha |0\rangle=0$ and $l_{i\bar{i}}, l_{\bar{i}3}, l_{3i}$ are positive integers. The normalisation factor in (\ref{psbasis}) is calculated in appendix \ref{app:on}.
 Above states can also be labelled by the eigenvalues of the number operators $\hat{N}_i=a^\dagger_i\cdot a_i, \hat{N}_2=a^\dagger_{\bar i}\cdot a_{\bar i}, \hat{N}_3=a^\dagger_3\cdot a_3$. I.e, \[|l_{i\bar{i}},l_{\bar{i}3},l_{3i}\rangle \equiv |n_i,n_{\bar i}, m\equiv n_3\rangle\] where $n_i=l_{i\bar{i}}+l_{3i}, n_{\bar i}=l_{i \bar{i}}+l_{\bar{i}3}, m\equiv n_3=l_{i3}+l_{\bar{i}3}$. 
The inverse relations are ${l}_{i\bar{i}}={n}_i+{n}_{\bar i}-{n}_3, {l}_{\bar{i}3}={n}_{\bar i}+{n}_3-{n}_i, {l}_{3i}={n}_3+{n}_i-{n}_{\bar i}$.
  Since, $E^2=\frac{\hat{N}}{2}(\frac{\hat{N}}{2}+1)$, number of prepotentials on a link is a measure of the electric flux through the link. The state $|n_i,n_{\bar i},m\rangle$ are precisely the gauge invariant coupled states obtained by the diagonalization procedure described in the introduction.The $n_i, m$ quantum numbers of the links meeting at a 3 vertex satisfies triangle inequalities. For eg, at the 3-vertex shown in figure (\ref{3vertex}) : 
\begin{align} |n_i-n_{\bar i}| \leq m\leq n_i+n_{\bar i}\label{trinq}\end{align}
In other words, $m(\vec{x})$ is limited by the value of $n_i(\vec{x})$ which in turn can take values freely between $0$ and $\infty$. Alternatively, $l_{i\bar{i}},l_{\bar{i}3},l_{3i}$ at a site can be varied independently.
The link constraint (\ref{u1prep}) can be translated in terms of the $l_{ij}$ quantum numbers. 
The physical state space can be described in terms of $l_{ij}$ at each site and the link constraints (\ref{u1prep}) on every link. This implies a collection of overlapping, closed electric flux loops on the ps-lattice. Equivalently, they can also be labeled by link variables $n_i$ and $m$ on the ps-lattice satisfying local triangle inequalities at each vertex. This equivalence can be realised only on the ps-lattice where the coordination number of any vertex is always three. 
\subsection{Hamiltonian on a ps-lattice}
\label{hamps}
The Hamiltonian on the ps-lattice in the prepotential representation is given by 
\begin{align}
H=\frac{{\tilde g}^2}{2}\sum\limits_{\vec{x}} \frac{\hat{N}_i(\vec{x})}{2}\Big(\frac{\hat{N}_i(\vec{x})}{2}+1\Big)+\frac{1}{2g^2}\sum\limits_{oct}\Big[2-TrU_o\Big]
\end{align}
Above, $\vec{x}$ denotes 3-vertices, $i=1,2$ according as the 3-vertex under consideration and $TrU_o=Tr(\hat{P}(31)_a\hat{P}(\bar{1}3)_b\hat{P}(32)_c\hat{P}(\bar{2}3)_d\hat{P}(3\bar{1})_e\hat{P}(13)_f\hat{P}(3\bar{2})_g\hat{P}(23)_h$ where $a,b,c,d,e,f,g,h$ are the eight 3-vertices (see figure \ref{pslattice}(b)) in the octagon and
{\footnotesize 
	\begin{align}
	&\hat{P}(ij)=
	\begin{pmatrix}
	\frac{1}{\sqrt{{\hat N}_i({\hat N}_j+1)}} \left(a^\dagger_i\cdot\tilde{a}^\dagger_j\right)&\frac{1}{\sqrt{{\hat N}_i({\hat N}_j+1)}}  \left(a^\dagger_i\cdot a_j\right)\\
	\frac{1}{\sqrt{({\hat N}_i+2)({\hat N}_j+1)}} \left(a_i\cdot a^\dagger_j\right)&\frac{1}{\sqrt{({\hat N}_i+2)({\hat N}_j+1)}} \left(\tilde{a}_i \cdot a_j\right)
	\end{pmatrix}
	\label{pijop}
	\end{align}
}
At each 3-vertex along the octagon, there are two links which lie on the octagon. These links are labelled by $i$ and $j$ in (\ref{pijop}). $\hat{P}(ij)_{11}/\hat{P}(ij)_{22}$ increases/decreases both $n_i,n_j$ by 1, $\hat{P}(ij)_{12}$ increases $n_i$ but decreases $n_j$ and  $\hat{P}(ij)_{21}$ increases $n_j$ but decreases $n_i$. $Tr{U}_o$ has $2^8$ terms. Each term changes $n_i/m$ at the eight links of the octagon by $\pm1$ thereby exhausting the $2^8$ possible ways to do it. The action of $TrU_o$ on the state $|n_i,n_{\bar i},m \rangle$ is illustrated in figure \ref{pslattice}(b).
This kind of construction can be done for any of the splitting schemes. We note that in \cite{rmi}, where figure \ref{ps}(a) type of splitting was envisaged and the corresponding hamiltonian was studied in the strong coupling expansion, it was found that it agrees with the naive strong coupling expansion. The matrix element of the Hamiltonian in the $|n_1,n_2,m\rangle$ basis is easily computed using (\ref{psbasis}) and (\ref{pijop}) [see eqn.(\ref{pijn})] and can be used to perform numerical analysis. 
\section{Gauge invariant path integral} 
\label{pisec}
The gauge invariant reformulation is described locally and completely in terms of $n_i,m$ fields without any redundant fields. However, the dynamics is still unconventionally complicated. To further simplify,
we introduce the phase operators \cite{caruthers} $e^{i\hat{\phi}_i},e^{i\hat{\chi}}$, conjugate to $n_i,m$. For this, we first extend the Hilbert space by increasing the domain of $n_i,m$ to $(-\infty,\infty)$. 
Phase operators are defined such that they satisfy the following commutation relations: 
\begin{align}
[{n}_i,e^{i\hat{\phi}_i}]=e^{i\hat{\phi}_i} \nonumber\\
[{m},e^{i\hat{\chi}}]=e^{i\hat{\chi}} 
\label{phase}
\end{align}
$e^{i\phi_i}/e^{-i\phi_i}, e^{i\chi}/e^{-i\chi}$ acts as step operators on the number states increasing/decreasing $n_i,m$ by 1.
We then define an eigenbasis of the phase operator $e^{i\hat{\phi}_i},e^{i\hat{\chi}}$:
\begin{align}
e^{i\hat{\phi}_i} |\phi_1,\phi_2,\chi\rangle = e^{i\phi_i} |\phi_1,\phi_2,\chi\rangle \nonumber\\
e^{i\hat{\chi}} |\phi_1,\phi_2,\chi\rangle = e^{i\chi} |\phi_1,\phi_2,\chi\rangle
\end{align}
The gauge invariant path integral is then constructed as the probability amplitude to go from the state $|\phi_1,\phi_2,\chi\rangle$ to $|\bar{\phi}_1,\bar{\phi}_2,\bar{\chi} \rangle$ in time t.
 The details of the construction is described in appendix \ref{app:pi}. The domain of $n_i,m$ is then restricted to positive values of $n_i,m$ within the path integral to get back to the original Hilbert space.  The Euclidean path integral thereby constructed is given by : 
\begin{widetext}
	\begin{align}
	Z=\int D\phi_1D\phi_2D\chi\sum\limits_{n_1,n_2,m} &e^{-\int dt\Bigg(\frac{\tilde{ g}^2}{2}\sum\limits_{\vec x}\Big(\frac{n_i}{2}\big[\frac{n_i}{2}+1\big]\Big)+\frac{1}{2g^2}\sum\limits_{oct}\Big[2- Tr\{P_{oct}\}\Big]+\sum\limits_{\vec x}\Big\{in_1\dot{\phi}_1+in_2\dot{\phi}_2+im\dot{\chi}+}\nonumber\\
		&{\hspace{1.5cm}}^{\lambda \Big(\Theta(-n_i)+\Theta(-m)+ \Theta\big((n_i-n_{\bar i})^2-m^2\big)+\Theta\big(m^2-(n_i+n_{\bar i})^2\big)\Big)\Big\}\Bigg)} 
	\label{pi}
	\end{align}
	Above, 
$P_{oct}=\bar{P}(31)_a {P}(\bar{1}3)_b \bar{P}(32)_c P(\bar{2} 3)_d$ $ \bar{P}(3\bar{1})_e {P}(13)_f \bar{P}(3{\bar 2})_g {P}(23)_h$ where $\bar{P}({ij})=\langle n_j,n_{\bar j},m|\hat{P}({ij})|\phi_j,\phi_{\bar j},\chi\rangle$ and ${P}({ij})=\langle\phi_j,\phi_{\bar j},\chi|\hat{P}({ij})|n_j,n_{\bar j},m\rangle$ given by,
	{\footnotesize 
		\begin{align}  
		P(ij)&=\begin{pmatrix}
		\sqrt{\frac{({n}_i+{n}_j+{n}_{\bar j}+4)({n}_i+{n}_j-{n}_{\bar j}+2)}{4({n}_i+1)({n}_j+2)}} e^{\frac{i}{2}({\phi}_i+{\phi}_j)} &\sqrt{\frac{(   {n}_i-{n}_j+{n}_{\bar j}+2)}{2({n}_i+1)({n}_j)}}  e^{\frac{i}{2}({\phi}_i-{\phi}_j)} \\
		\sqrt{\frac{({n}_j-{n}_i+{n}_{\bar j}+2)}{2({n}_i+1)({n}_j+2)}} e^{-\frac{i}{2}({\phi}_i-{\phi}_j)}&\sqrt{\frac{({n}_i+{n}_j+{n}_{\bar j}+2)({n}_i+{n}_j-{n}_{\bar j})}{4({n}_i+1)({n}_j)}}
		e^{-\frac{i}{2}({\phi}_i+{\phi}_j)}
		\end{pmatrix}~~~~;~n_j\neq0, n_i\neq 0
		\nonumber\\
		{\bar P}(ij)&= 
		\begin{pmatrix}
		\sqrt{\frac{({n}_i+{n}_j+{n}_{\bar j}+2)({n}_i+{n}_j-{n}_{\bar j})}{4({n}_i)({n}_j+1)}} e^{\frac{i}{2}({\phi}_i+{\phi}_j)} &\sqrt{\frac{({n}_i-{n}_j+{n}_{\bar j})}{2({n}_i)({n}_j+1)}}  e^{\frac{i}{2}({\phi}_i-{\phi}_j)}\\
		\sqrt{\frac{({n}_j-{n}_i+{n}_{\bar j})}{2({n}_i+2)({n}_j+1)}} e^{-\frac{i}{2}({\phi}_i-{\phi}_j)}&\sqrt{\frac{({n}_i+{n}_j+{n}_{\bar j}+4)({n}_i+{n}_j-{n}_{\bar j}+2)}{4({n}_i+2)({n}_j+1)}}
		e^{-\frac{i}{2}({\phi}_i+{\phi}_j)}
		\end{pmatrix} ~~~~;~n_i\neq0,n_j\neq0
		\label{pijmain}
		\end{align}
	}
Equation (\ref{pijop}) implies, when $n_j=0$, 
the second column of $P({ij})$ is 0 and when $n_i=0$, second row of $P(ij)$ is 0. Similarly,  when $n_i=0$ the first row of $\bar{P}({ij})$ is 0 and when $n_j=0$ the first column is 0. There is a matrix $P(ij)$ and ${\bar P}({ij})$ associated with each 3-vertex along an octagon under consideration on the split lattice. Our convention is, at a 3 vertex where the links $i,j,{\bar j}$ meet, $i,j$ are the links in the direction of the octagon and ${\bar j}$ is the spectator. In (\ref{pi}), the step function $\Theta$ is used to implement the positivity condition of $n_i,m$ as well as the triangle inequality of $n_i,m$ at each 3-vertex by taking $\lambda \rightarrow \infty$. $i=1,2$ in $n_i$ according to the 3-vertex under consideration. $n_i,m$ and $\phi_i, \chi$ fields are functions of time t and lattice points. 
\end{widetext}
\section{Weak coupling, continuum limit}
\label{weakc}
 The deduction in the previous sections has been exact. Now we turn to making a useful ansatz which allows us to make weak coupling expansion. We remind ourselves that the magnetic term is strictly positive and its lowest value occurs when $U(\vec{x})$ reach $\mathbf{1}$, this in the gauge invariant configuration space leads to $U_o \rightarrow 1$ and $P({ij}), {\bar P}({ij}) \rightarrow 1$. Furthermore, all the integer fields $n_i,m$ are always positive and hence $\langle n_i \rangle$ and $\langle m \rangle$ are non zero. If we assume $\langle n_i \rangle =N$ and insist that $P({ij}), {\bar P}({ij}) \sim 1$, From eqn.(\ref{pijmain}), we note that $\langle m \rangle=2N$ and N is large with $\phi_i,\chi \rightarrow 0$, as the only solution. This mean field ansatz is indeed possible only in the point splitting figure \ref{ps}(c). In the other possible schemes, such a mean field with corresponding $P({ij}),{\bar P}({ij}) \sim 1$ is not possible.  
We rescale $\phi_i,\chi \rightarrow g\phi_i,g\chi$ and make the 
 substitution, $n_i\equiv N+\tilde{n}_i,m\equiv 2N+\tilde{m}$. 
 $TrP_{oct}$ consists of $2^8$ terms corresponding to all possible fluctuations generated by the plaquette term along the octagon(see figure \ref{pslattice}(b)). In each of the terms, the off diagonal terms come in pairs. This is due to the link constraint (\ref{u1}). The off-diagonal terms of $P(ij)$ and $\bar{P}(ij)$ are at least of the order of $\frac{1}{\sqrt{N}}$. Therefore, the terms can be classified according to the number of off diagonal terms occurring. Such a classification corresponds to an expansion in $\frac{1}{N}$. The leading term $\frac{1}{N^0}$ is the term which does not contain any off diagonal terms and corresponds to the term which either increases(+) or decreases(-) the field $n_i, m$ at every link along the octagon. Every other term consists of several flips $+\rightarrow-$ or $-\rightarrow +$ along the octagon and each flip brings in a $\frac{1}{\sqrt{N}}$. A consistent diagram necessarily has even number of flips only. The second leading term has two flips and hence is of the order of $\frac{1}{N}$. 
 The next term $\frac{1}{N^2}$ consists of four such flips. After a straight forward tedious calculation, keeping the terms up to the order $\frac{1}{N}$ and quadratic in $\phi_i, \chi$, we get 
\begin{widetext} 
\begin{align}
TrU_o\approx 2- \Big(\frac{1}{4N^2}{\tilde m}^2 +V(\phi_1,\phi_2,\chi)\Big)
\label{truo}
\end{align}
\begin{align}
V(\phi_1,\phi_2,\chi)=\frac{g^2}{2}\Bigg\{&\Big[(\Delta_1\big(\phi_2-\frac{1}{2}\Delta_2\chi\big)-\Delta_2\big(\phi_1+\frac{1}{2}\Delta_1\chi)\big)\Big]^2+ \frac{4}{N}\bigg[16\Big[(\phi_1+\frac{1}{2}\Delta_1\chi)^2+(\phi_2-\frac{1}{2}\Delta_2\chi)^2+\chi^2\Big]\nonumber\\
&-\Big[\Delta_1\big(\phi_2-\frac{1}{2}\Delta_2\chi\big)-\Delta_2\big(\phi_1+\frac{1}{2}\Delta_1\chi\big)+\Delta_1\Delta_2\chi\Big]^2-(\Delta_1\Delta_2\chi)^2\bigg]\Bigg\}
\end{align}
\end{widetext}
In the above expression, $\Delta_i$ is the forward difference operator defined by $\Delta_i f(\vec{x})=f(\vec{x}+\hat{i})-f(\vec{x})$. 
The details of the above weak coupling expansion is described in appendix \ref{wc}. The path integral becomes: 
\begin{widetext}
\begin{align}
Z=\int\limits_{-\frac{\pi}{g}}^{\frac{\pi}{g}} D\phi_1D\phi_2D\chi  \sum_{\substack{{\tilde n}_1=-N\\{\tilde n}_2=-N}}^{\infty}\sum_{\tilde m=-M_1}^{M_2}e^{-\int dt \sum\limits_{sites}\Bigg[ \frac{\tilde{g}^2}{8}\Big( {\tilde n}_1^2+{\tilde n}_2^2+2N^2\Big)+\frac{1}{8g^2N^2}{\tilde m}^2 +i{\tilde n}_ig\dot{\phi}_i+i{\tilde m} g\dot{\chi}+\frac{1}{2g^2}V(\phi_1,\phi_2,\chi)]\Bigg]}
\end{align}
\end{widetext}
Above, $M_1= min(\tilde{n}_1+\tilde{n}_{\bar 1},\tilde{n}_2+\tilde{n}_{\bar 2}), M_2=max(2N-|\tilde{n}_1-\tilde{n}_{\bar 1}|,2N-|\tilde{n}_2-\tilde{n}_{\bar 2}|)$.
This is due to the fact that $\tilde{m}$ is limited by the triangle inequality (\ref{trinq}). In the weak coupling limit, $M_1\approx0, M_2\approx 2N$. Therefore, triangle inequality becomes irrelevant upto the leading term in the weak coupling limit and theta function terms used to implement the triangle inequality and the positivity condition can be removed. 
Using Euler-Maclaurin formula to convert the summation over $\tilde{n}$ and ${\tilde m}$ to integration and 
performing the integral, we get upto an irrelevant constant C:
\begin{widetext}
\begin{align}
&Z=C\int D\phi_1D\phi_2D\chi  e^{-W[N,\phi_i,\dot{\phi}_i,\chi,\dot{\chi}]}
\end{align}
{\footnotesize 
	\begin{align}
	W[N,\phi_i,\dot{\phi}_i,\chi,\dot{\chi}]=\int dt \sum\limits_{sites}\Bigg[
	2g^2\frac{\dot{\phi_1}^2+\dot{\phi_2}^2}{\tilde{g}^2}+g^2\frac{\dot{\chi}^2}{1/(2g^2N^2)}+\frac{1}{2g^2}V(\phi_1,\phi_2,\chi)\Bigg]
	\end{align}
}
\end{widetext}
We now make the following transformation, 
\begin{align}
\phi_i=\frac{1}{\sqrt{-\Delta^2}}(\Delta_i\eta+\epsilon_{ij}\delta_j\psi)+\frac{1}{2}\epsilon_{ji}\Delta_j \chi
\label{trans}
\end{align}
In (\ref{trans}), $\Delta^2$ is the lattice laplacian and $\delta_j$ is the backward difference operator defined as $\Delta^2 =\sum_i (\Delta_i-\delta_i)$
; $\delta_j f(\vec{x})=f(\vec{x})-f(\vec{x}-\hat{j})$.
Ignoring constant terms, the path integral becomes: 
\begin{widetext}
{\footnotesize 
\begin{align}
&Z= \int D\psi D\eta D\chi~  e^{-\int dt \sum\limits_{sites}\Big[\frac{2g^2}{\tilde{g}^2}\big(\dot{\eta}^2+\dot{\psi}^2\big)+2g^4N^2\dot{\chi}^2+\frac{2g^2}{\tilde{g}^2}\Big( \frac{1}{4}\big((\Delta_1 \dot{\chi})^2+ (\Delta_2\dot{\chi})^2\big)+\dot{\chi}\frac{1}{\sqrt{-\Delta^2}}\big[(\delta_1\Delta_1-\delta_2\Delta_2)\dot{\eta}+2\delta_1\delta_2\dot{\psi}\big]\Big)+\frac{1}{2g^2}V(\psi,\eta,\chi)\Big]}
\label{pif}
\end{align}
\begin{align}
&V(\psi,\eta,\chi)=	2g^2\bigg\{\frac{1}{4} (\Delta\psi)^2+\frac{1}{N} \bigg[16(\eta^2+\psi^2+\chi^2)-(\Delta\psi)^2-2(\Delta_1\Delta_2 \chi)^2+\Delta^2 \chi\frac{\Delta_1\Delta_2}{{\sqrt{-\Delta^2}}} \chi\bigg]\bigg\}
\end{align}
}
\end{widetext}
Above, $(\Delta \psi)^2\equiv (\Delta_1 \psi)^2+(\Delta_2 \psi)^2$. 
In order to cast the $\psi$ terms in the canonical form, we rescale $\psi\rightarrow\sqrt{2} \psi$. Now, velocity of light being 1 requires
\begin{align}
\frac{g^2}{\tilde{g}^2}=\frac{1}{8}a^2
\end{align} 
The continuum limit is now taken by making 
\begin{align}
\frac{64}{N}=M^2a^2
\end{align}
where $M$ is the mass in the continuum and $a$ is the lattice constant. Since, $g^2=a, \tilde{g}^2=\frac{8}{a}$; $N=\frac{64}{M^2g^4}=\frac{\tilde{g}^4}{M^2} $.
Consequently,
 the euclidean inverse propagators in the energy-momentum space to the leading order are 
\begin{align}
\psi &: p_0^2+M^2+\vec{p}^2+O(a^2)\nonumber\\
\eta &: p_0^2+M^2+O(a^4)\nonumber\\
\chi&: \#a^2\vec{p}^2p_0^2+M^2+O(a^4)
\label{disper}
\end{align}
Above, $\#$ denotes a real positive constant.
Therefore, $\psi$ is a relativistic particle with mass $M$ 
 and $\chi$ do not propagate.  $\eta$ may propagate due to higher order corrections.
\section{Summary and Discussion}    
A complete gauge invariant Hilbert space could be constructed in the electric space \cite{ani} by solving the local Gauss law at each site on the lattice. This leads to complicated dynamics as the matrix element of Hamiltonian in this basis is given by higher Wigner coefficients \cite{ani,prep}. In order to construct a gauge invariant basis where the dynamics is more transparent, it is convenient to break down the link operators into simpler objects which transform under the fundamental representation of the gauge group namely, the Schwinger Bosons. This leads to the prepotential representation. However, the local gauge invariant objects constructed at each site by contracting various Schwinger bosons are not all independent. This is the analogue in electric space of the Mandelstam constraints \cite{loop2} in the Wilson loop space. In order to construct an independent basis, we point split each site of the lattice. 

Advantages of point splitting are many fold. Point splitting allow us to construct a local set of independent operators. This in turn allow us to construct a complete, local, orthonormal, gauge invariant basis (\ref{psbasis}) of SU(2) LGT.  As a consequence, the Mandelstam constraints mentioned above is completely bypassed.
Secondly, on a split lattice, the matrix elements of $TrU_p$ (see (\ref{pijn})) becomes much simpler to analyze, especially in the weak coupling limit. At each split site (3-vertex), the matrix element becomes a simple algebraic expression (see (\ref{pijn})) involving $n_i,m$ fields, as opposed to the 6j coefficients on the unsplit lattice. Therefore, we are in effect writing down 6j symbols as the product of 2 simpler  gauge invariant expressions which can be thought of as the  gauge invariant projection of standard 3j symbols.
Further, the dynamics is local and the exact matrix elements in the ps-basis (\ref{psbasis}) is easily written down (see (\ref{pijn})). These
matrix elements could be much more convenient in a numerical analysis as opposed to the higher Wigner coefficients \cite{ani,prep} which occur on the unsplit lattice. More importantly, point splitting enables us to analyze a theory of interacting closed loops in terms of 
local quantum numbers satisfying triangle inequalities. 

     There doesn't seem to be any serious hindrance in the construction of a point splitting scheme in higher dimensions and SU(N)\cite{prep} group. 
 Construction of a complete local gauge invariant basis of the Hilbert space similar to what is described in this paper can be envisaged for these cases as well. 
    
      On inclusion of fermions, the local Gauss law is modified at each site. Therefore, a modified point splitting scheme involving fermions has to be constructed which leads to a new complete, gauge invariant, local ps-basis. 
    Such a construction leads to new singlet operators involving fermions along with the usual singlets involving only Schwinger bosons.

In the weak coupling $g\rightarrow 0$ limit, $U\rightarrow {\bf 1}$. In section \ref{weakc}, we found that writing the fields $n_i, m$ around a large mean value $N, 2N$ and small $\phi, \chi$, makes $U \rightarrow {\bf 1}$. This ansatz is also consistent with the triangle inequality at each 3-vertex on the ps-lattice. 
 In the weak coupling limit, such a vacuum is dominated by small electric flux loops in space carrying large fluxes.
In other schemes, there is no mean field ansatz consistent with triangle inequalities at each 3-vertex which takes $U\rightarrow {\bf 1}$. Therefore, even though other splitting schemes leads to different basis of the same Hilbert space and are therefore completely equivalent in an exact analysis, $g\rightarrow 0$ can not be achieved by a simple mean field ansatz.
 However, in other splitting schemes, mean field ansatz can be constructed where $U\rightarrow c {\bf 1}$, where $c<1$ . Such an ansatz is suitable to study $g$ small but not 0 . It will be interesting to see whether such schemes lead to a Lorentz covariant fixed point at a finite but small value of $g$. 

  The dispersion relations (\ref{disper}) were deduced analytically on the lattice upto lowest order in the weak coupling expansion. It is not evident that everything is Lorentz covariant. $\psi$ mode turns out to be a relativistic scalar particle satisfying relativistic dispersion relation and $\chi$ is a massive mode (on the lattice) which does not survive the continuum limit. $\eta$ could become a propagating mode due to higher order terms. The lowest excited state $\psi$ being non-degenerate is consistent with the results from the existing literature \cite{nair,loop3}. In order for the mass gap to be compared, it has to be written in terms of string tension, i.e, we need to introduce heavy SU(2) charged fermions. 
  As described earlier inclusion of fermions changes the Gauss law at sites and hence the point splitting scheme and the construction of the gauge invariant Hilbert space.  
  Interaction terms in the higher orders in the weak coupling expansion have to be investigated. 

\acknowledgments

\noindent {\it Acknowledgments: We thank  Manu Mathur for useful discussions.}

\appendix
\section{Normalization of local basis states at a 3-vertex}
\label{app:on}
A general gauge invariant state at a 3 vertex shown in figure (\ref{3vertex}) is : 
{\footnotesize
	\begin{align}
	|l_{i\bar{i}},l_{\bar{i}3},l_{3i}\rangle= (a^\dagger_i \cdot \tilde{a}^\dagger_{\bar{i}})^{l_{i\bar{i}}} (a^\dagger_{\bar{i}} \cdot \tilde{a}^\dagger_3)^{l_{\bar{i}3}} (a^\dagger_3 \cdot \tilde{a}^\dagger_i)^{l_{3i}}~ ~ |0\rangle
	\end{align}
}
The inner product between two such states is given by: 
\begin{align}
\label{step1}
&\langle l_{i{\bar i}},l_{i3},l_{{\bar i}3}|l'_{i{\bar i}},l'_{i3},l'_{{\bar i}3} \rangle = \langle 0|(a_{\bar i}\cdot\tilde{a}_3)^{l_{{\bar i}3}}~(a_i\cdot\tilde{a}_3)^{l_{i3}}~\nonumber \\
&(a_i\cdot\tilde{a}_{\bar i})^{l_{i{\bar i}}}(a^\dagger_i \cdot \tilde{a}^\dagger_{\bar i})^{l'_{i{\bar i}}}  (a^\dagger_i \cdot \tilde{a}^\dagger_3)^{l'_{i3}}~(a^\dagger_{\bar i} \cdot \tilde{a}^\dagger_3)^{l'_{{\bar i}3}} |0\rangle
\end{align}
Using the relation
\begin{align}
\label{comm1}
\Big[a_n\cdot {\tilde a}_{m}~,~a^\dagger_n\cdot \tilde{a}^\dagger_{m}\Big]= \hat{N}_{\bar n}+\hat{N}_m+2
\end{align}
repeatedly to shift $(a_i\cdot\tilde{a}_{\bar i}),(a_i\cdot\tilde{a}_{3})$ and $(a_{\bar{i}}\cdot\tilde{a}_{3})$ to the right, we get 
\begin{align}
\langle l_{i{\bar i}},l_{i3},l_{{\bar i}3}|l'_{i{\bar i}},l'_{i3},l'_{{\bar i}3} \rangle 
&=\delta_{l_{i{\bar i}}~l'_{i{\bar i}}}\delta_{l_{i3}~l'_{i3}}\delta_{l_{{\bar i}3}~l'_{{\bar i}3}}~\nonumber \\&(l_{i{\bar i}}+l_{i3}+l_{{\bar i}3}+1)!(l_{i{\bar i}})!(l_{i3})!(l_{{\bar i}3})!
\end{align}

Therefore, the normalized states are :
{\footnotesize 
\begin{align}
|l_{i{\bar i}},l_{i3},l_{{\bar i}3}\rangle = \frac{(a^\dagger_i \cdot \tilde{a}^\dagger_{\bar i})^{l_{i{\bar i}}} ~ (a^\dagger_i \cdot \tilde{a}^\dagger_3)^{l_{i3}}~(a^\dagger_{\bar i} \cdot \tilde{a}^\dagger_3)^{l_{{\bar i}3}}}{\sqrt{(l_{i{\bar i}}+l_{i3}+l_{{\bar i}3}+1)!(l_{i{\bar i}})!(l_{i3})!(l_{{\bar i}3})!}} ~ |0\rangle
\label{onstate}
\end{align}
}
A couple of comments are in order here. In general, 
we can associate a sign \cite{rmi} for the states. However, in our calculation as long as the same sign convention is applied to both the bras and kets, all our expectation values are independent of the sign. States defined by (\ref{onstate}), are a special case of what occurs in the context of addition of angular momentum. In the latter, normalization depends on the azimuthal quantum numbers as well. Because of local gauge invariance, in our context (\ref{onstate}), they depend only on the casimirs.
\section{Path integral formulation}
\label{app:pi}
The commutation relations (\ref{phase}) implies that $e^{i\phi_i},e^{i\chi}$ increases the ${n}_i,{m}$ quantum numbers respectively by one. Hence, when $\hat{P}(ij)$ acts on the number state, $(a^\dagger_i\cdot \tilde{a}^\dagger_j), (a^\dagger_i\cdot {a}_j), (a_i\cdot {a}^\dagger_j), ( \tilde{a}_i\cdot {a}^\dagger_j)$ can be replaced by $n$ dependent factors times $e^{\frac{i}{2}(\phi_i+\phi_j)},e^{\frac{i}{2}(\phi_i-\phi_j)},e^{\frac{-i}{2}(\phi_i-\phi_j)},e^{\frac{-i}{2}(\phi_i+\phi_j)}$. The factor of $\frac{1}{2}$ comes from the fact that each $n$ is shared by two 3-vertices. Therefore, for $n_j\neq 0, n_i\neq 0$, 
\begin{widetext}
	\begin{align}
	\hat{P}(ij)|n_j,n_{\bar{j}},m\rangle 
	&=
	\begin{pmatrix}
e^{\frac{i}{2}(\hat{\chi}+\hat{\phi}_j)}	\sqrt{\frac{(\hat{N}_{\bar j}+\hat{N}_j+\hat{N}_i+4)(\hat{N}_j-\hat{N}_{\bar j}+\hat{N}_i+2)}{4(\hat{N}_i+1)(\hat{N}_j+2)}}  &e^{\frac{i}{2}(\hat{\chi}-\hat{\phi}_j)} \sqrt{\frac{(\hat{N}_{\bar j}-\hat{N}_j+\hat{N}_i+2)}{2(\hat{N}_i+1)(\hat{N}_j)}}  \\
e^{-\frac{i}{2}(\hat{\chi}-\hat{\phi}_j)}	\sqrt{\frac{(\hat{N}_{\bar j}+\hat{N}_j-\hat{N}_i+2)}{2(\hat{N}_i+1)(\hat{N}_j+2)}} &e^{-\frac{i}{2}(\hat{\chi}+\hat{\phi}_j)}\sqrt{\frac{(\hat{N}_{\bar j}+\hat{N}_j+\hat{N}_i+2)(\hat{N}_j-\hat{N}_{\bar j}+\hat{N}_i)}{4(\hat{N}_i+1)(\hat{N}_j)}}	
	\end{pmatrix} 
	|n_j,n_{\bar{j}},m\rangle 
\label{pijn}
	\end{align}
	 From (\ref{pijop}), for $n_j=0$, the second column of $\hat{P}({ij})$ becomes 0 and for $n_i=0$, the second row becomes zero.
	
Path integral is given by: 
	\begin{align}
Z=\langle\bar{\phi}_1,\bar{\phi}_2,\bar{\chi}~|~e^{-Ht}~|\phi_1,\phi_2,\chi~\rangle = \prod_{i=1}^{n-1}\int d\phi_{1i}d\phi_{2i}d\chi_{i}& ~\bigg\langle \phi_{1i+1},\phi_{2i+1},\chi_{i+1}\bigg|e^{-H\delta t}\bigg| \phi_{1i},\phi_{2i},\chi_{i}\bigg\rangle
\nonumber\\ 
&e^{-\lambda \Big(\Theta(-n_i)+\Theta(-m)+ \Theta\Big(-\big(m^2-(n_i-n_{\bar i})^2\big)+\Theta\big((n_i+n_{\bar i})^2-m^2\big)\Big)} 
\label{pitemp}
	\end{align}
	Above, $t=(n+1)\delta t$ and $\lambda$ is a real parameter. The step function $\Theta$ is used to restrict $n_i,m$ to positive values satisfying triangle inequality by taking the limit $\lambda \rightarrow \infty$. 
	Now,
	{\footnotesize 
		\begin{align}
		\label{1slice}
		&\bigg\langle \phi_{1i+1},\phi_{2i+1},\phi_{3i+1}\big|e^{-H\delta t}\big| \phi_{1i},\phi_{2i},\phi_{3i}\big\rangle = \sum_{n_1,n_2,n_3}\big\langle \phi_{1i+1},\phi_{2i+1},\phi_{3i+1}\big|e^{-H\delta t}\big|n_1,n_2,n_3\big\rangle \big\langle n_1,n_2,n_3\big| \phi_{1i},\phi_{2i},\phi_{3i}\big\rangle \nonumber\\
		&=\sum\limits_{n_1,n_2,n_3}e^{-\delta t \bigg[\sum\limits_{s}\big[i({ n}_{1}\dot{\phi}_{1i}+{n}_2\dot{\phi}_{2i}+{m}\dot{\chi}_{i})+\frac{\tilde{g}^2}{2}\big(n^2_1(s)+n^2_2(s)\big)\big]+\frac{1}{2g^2} \sum\limits_{oct}\big[2-Tr P_{oct}\big] \bigg]}
		\end{align}
	}
	Above, $\dot{\phi}_{1i} = \frac{\big(\phi_{1i+1}-\phi_{1i}\big)}{\delta t}~,~\dot{\phi}_{2i} = \frac{\big(\phi_{2i+1}-\phi_{2i}\big)}{\delta t}~,~ \dot{\phi}_{3i} = \frac{\big(\phi_{3i+1}-\phi_{3i}\big)}{\delta t}$ , 
$P_{oct}=\bar{P}(31)_a P(\bar{1}3)_b \bar{P}(32)_c P(\bar{2} 3)_d$ $ \bar{P}(3\bar{1})_e {P}(13)_f \bar{P}(3{\bar 2})_g {P}(23)_h$ where,  $\bar{P}({ij})=\langle n_j,n_{\bar j},m|\hat{P}({ij})|\phi_j,\phi_{\bar j},\chi\rangle$ and ${P}({ij})=\langle\phi_j,\phi_{\bar j},\chi|\hat{P}({ij})|n_j,n_{\bar j},m\rangle$ given by (\ref{pijmain}).
Plugging (\ref{1slice}) back into (\ref{pitemp}) and taking the limit $\delta t \rightarrow 0$ gives the path integral (\ref{pi}).We have used the $\lambda$ term in (\ref{pitemp}) to avoid some technical difficulties alluded to in ref \cite{caruthers}. In our context, in the path integral all the $n_i,m$ fields are restricted to be positive and satisfy local triangle inequality as expected.
	\end{widetext}
\section{Weak coupling expansion of $TrU_o$}
\label{wc}
As described in section \ref{weakc}, in the continuum limit $\phi_i,\chi$ becomes small and $N$ becomes large. This motivates us to rescale $\phi_i,\chi \rightarrow g\phi_i,g\chi$ and make the substitution, $n_i=N+{\tilde n}_i,m=2N+{\tilde m}$. This allows us to expand $TrU_o$ as powers of the small parameter $\frac{1}{N}$.  We write the $n$ dependent terms $f(n)$ in $P(ij)$ and $\bar{P}{ij}$ as $ e^{ln (f(n))}$
and expand the $ln$ in powers of $\frac{1}{N}$. This gives for eg: 
{\footnotesize 
\begin{widetext}
	\begin{align}  
P_{11}(31)&=
 e^{\frac{{\tilde n}_{\bar 1}+{\tilde n}_1+{\tilde n}_3+4}{8N}-\frac{1}{4}\Big(\frac{{\tilde n}_{\bar 1}+{\tilde n}_1+{\tilde n}_3+4}{4N}\Big)^2+\frac{{\tilde n}_1-{\tilde n}_{\bar 1}+{\tilde n}_3+2}{(4N)}-\frac{1}{4} \Big(\frac{{\tilde n}_1-{\tilde n}_{\bar 1}+{\tilde n}_3+2}{2N}\Big)^2-\frac{{\tilde n}_3+1}{4N}+\frac{1}{4}\Big(\frac{{\tilde n}_3+1}{2N}\Big)^2-\frac{{\tilde n}_1+2}{2N}+\frac{1}{4}\Big(\frac{{\tilde n}_1+2}{N}\Big)^2+\frac{i}{2}({\phi}_3+{\phi}_1)} \nonumber\\
P_{12}(31)&=\frac{1}{\sqrt{N}} e^{\frac{{\tilde n}_{\bar 1}-{\tilde n}_1+{\tilde n}_3+2}{4N}-\frac{1}{4}\Big(\frac{{\tilde n}_{\bar 1}-{\tilde n}_1+{\tilde n}_3+2}{2N}\Big)^2-\frac{{\tilde n}_3+1}{4N}+\frac{1}{4}\Big(\frac{{\tilde n}_3+1}{2N}\Big)^2-\frac{{\tilde n}_1}{2N}+\frac{1}{4}\Big(\frac{{\tilde n}_1}{N}\Big)^2+\frac{i}{2}({\phi}_3-{\phi}_1)} \nonumber \\
P_{21}(31)&=\frac{1}{\sqrt{N}} e^{\frac{{\tilde n}_{\bar 1}+{\tilde n}_1-{\tilde n}_3+2}{8N}-\frac{1}{2}-\frac{1}{4}\Big(\frac{{\tilde n}_{\bar 1}+{\tilde n}_1-{\tilde n}_3+2}{4N}-1\Big)^2-\frac{{\tilde n}_3+1}{4N}+\frac{1}{4}\Big(\frac{{\tilde n}_3+1}{2N}\Big)^2-\frac{{\tilde n}_1+2}{2N}+\frac{1}{4}\Big(\frac{{\tilde n}_1+2}{N}\Big)^2-\frac{i}{2}({\phi}_3-{\phi}_1)} \nonumber \\
P_{22}(31)&=
e^{\frac{{\tilde n}_{\bar 1}+{\tilde n}_1+{\tilde n}_3+2}{8N}-\frac{1}{4}\Big(\frac{{\tilde n}_{\bar 1}+{\tilde n}_1+{\tilde n}_3+2}{4N}\Big)^2+\frac{{\tilde n}_1-{\tilde n}_{\bar 1}+{\tilde n}_3}{(4N)}-\frac{1}{4} \Big(\frac{{\tilde n}_1-{\tilde n}_{\bar 1}+{\tilde n}_3}{2N}\Big)^2-\frac{{\tilde n}_3+1}{4N}+\frac{1}{4}\Big(\frac{{\tilde n}_3+1}{2N}\Big)^2-\frac{{\tilde n}_1}{2N}+\frac{1}{4}\Big(\frac{{\tilde n}_1}{N}\Big)^2-\frac{i}{2}({\phi}_3+{\phi}_1)} 
\label{pijapprox}
\end{align}
\end{widetext}
}
and similar expressions for $P(ij)$ and ${\bar P}(ij)
$. We now make the following field redefinition: 
\begin{align}
n_1(\vec{x})/\phi_1(\vec{x}) \rightarrow (-1)^{x_1+x_2} n_1(\vec{x})/\phi_1(\vec{x}) \nonumber \\
n_2(\vec{x})/\phi_2(\vec{x}) \rightarrow (-1)^{x_1+x_2+1} n_2(\vec{x})/\phi_2(\vec{x})\nonumber\\
m(\vec{x})/\chi(\vec{x}) \rightarrow (-1)^{x_1+x_2+1} m(\vec{x})/\chi(\vec{x})
\end{align}
Above redefinition implies 
$P(ij)\approx{\bar P}(ij)$. Expanding the exponential in $TrU_o$ and keeping terms upto $O(1/N^2)$ and $O(g^2)$, we arrive at eqn.(\ref{truo}). 

\end{document}